\documentclass[epj]{svjour}
\usepackage{graphics}
\begin{document}
\title{Overlap of electron shells in $\beta$ and double-$\beta$ decays}
\author{M.~I.~Krivoruchenko\inst{1} \and K.~S.~Tyrin\inst{2}
}                     
%
%
\institute{Institute for Theoretical and Experimental Physics\\ National Research Center "Kurchatov Institute"\\ B. Cheremushkinskaya 25 117218 Moscow, Russia \and National Research Center "Kurchatov Institute"\\ pl. Kurchatova 1 123182 Moscow, Russia}
\date{Received: date / Revised version: date}
%
\abstract{
The $\beta$ and double-$\beta$ decay channels, which are not accompanied by excitation of the electron shells,
are suppressed due to the nonorthogonality of the electron wave functions of the parent and daughter atoms.
The effect is sensitive to the contribution of the outer electron shells.
Since valence electrons participate in chemical bonding and collectivize in metals,
the decay rates of the unstable nuclides
are modified when they are embedded in a host material.
Core electrons are less affected by
the environment, and their overlap amplitudes are more stable. The suppression effect is
estimated for $ \beta^- $ decay of $^{87}$Kr, electron capture in $^{163}$Ho, and $2\beta^-$ decays of $^{76}$Ge,
$^{100}$Mo, $^{130}$Te, and $^{136}$Xe. The overlap amplitude of the electron shells enters the relationship between
the half-life of neutrinoless $2\beta$ decay and the effective electron neutrino Majorana mass.
} 
\maketitle
%
\section{Introduction}

Neutrinos are likely the most promising particles  in the search for new physics beyond the Standard Model. Their electroneutrality and exceptionally low masses  raise the question of whether neutrinos are Majorana particles. Majorana neutrinos do not support conservation of the total lepton number.
The nonconservation of the total lepton number is sought in the processes of neutrinoless $ 2\beta^- $ decay, neutrinoless double-electron capture (2EC), and others. In the quark sector of the Standard Model, processes with nonconservation of the baryon number, such as proton decay and neutron-antineutron oscillations, play a similar fundamental role. The conservation laws of the total lepton and baryon numbers are not supported by local gauge symmetries of the Standard Model, and in a more general context, these conservation laws can be violated.

The effective electron neutrino Majorana mass, $m_{\beta\beta}$, can be extracted from the measurement of the half-life of
the neutrinoless $ 2 \beta^- $ decay and 2EC
in the case that the mechanism leading to these processes is the exchange of a light Majorana neutrino.
The amplitudes of the neutrinoless $ 2 \beta^- $ decay and 2EC processes,
which are not accompanied by excitation of the electron shells, are proportional
to $m_{\beta\beta}$, the nuclear matrix element,
the axial-vector coupling constant $g_A$, and the overlap of the electron wave functions
of the parent and daughter atoms. The effective neutrino masses
that can potentially be extracted from the experiment
depend on the overlap amplitude of the electron shells.

In this paper, we discuss the overlap effect in decays, accompanied by a change in the electric charge of the nucleus.
In the next section, the overlap amplitude of the electron levels with identical quantum numbers
and the overlap amplitude of the electron shells are analytically found in a simple nonrelativistic model.
The model is then generalized by taking into account shielding of the nuclear charge caused by the inner electrons.
The estimates show the dominance of the contribution of the electrons in outer orbits to the overlap amplitude of the electron shells.
We also consider the relativistic shell model based on the Dirac equation,
in which the effective charge of the nucleus is self-consistently determined using the semi-empirical data on the electron binding energies at individual orbits.
There are three possible $\beta$ decay processes: $\beta^-$, $\beta^+$, and electron capture (EC),
and four possible  double-$\beta$ decay processes: $2\beta^-$, $2\beta^+$, $\beta^+$EC, and 2EC.
In Sect. 3, numerical results based on the relativistic shell model are presented
for
$\beta^-$ decay of $^{87}$Kr;
EC in $^{163}$Ho, studied in the ECHo experiment in order to determine the absolute scale of neutrino masses \cite{Gastaldo:2013wha,Hassel:2016ixd};
$2\beta^-$ decays of $^{76}$Ge, $^{130}$Te and $^{136}$Xe,
the neutrinoless mode of which is searched for by the collaborations
GERDA \cite{Agostini:2018tnm},
CUORE \cite{CUORE2015},
and KamLAND-Zen \cite{KamLAND-Zen:2016pfg}, respectively;
and neutrinoless $2\beta^-$ decay of $^{100}$Mo, searched for by the collaborations
NEMO-3 \cite{NEMO32015},
CUPID-Mo \cite{Armengaud:2019loe}, and
AMoRE \cite{Alenkov:2019jis}.
Finally, we discuss modification of the total $\beta$ and $2\beta$ decay probabilities caused by
the overlap of the electron shells.

\section{Overlap amplitude}

The $\beta$ and double-$\beta$ decays are accompanied by a change in the nuclear charge, $Z$, by one or two units, respectively.
Electrons, which are initially in the stationary states of the parent atom,
turn into a superposition of stationary states of the daughter atom.
For example, the nonorthogonality of the electron wave functions leads to modification of the energy spectrum
of EC in $^{163}$Ho \cite{AmandFaessler2015,Faessler:2016hxd,Brass2018}.
In this section, we estimate the survival probability for the ground state of the electron shells
in the $\beta$ and double-$\beta$ decays.
A similar effect of overlapping wave functions of nucleons in double-$\beta$ decays
is discussed in Ref.~\cite{Simkovic2003}.

\subsection{Analytic approach}

The standard separation of variables in the energy eigenfunctions of the nonrelativistic Coulomb problem gives
\begin{equation}
\Psi _{nlm} (\mathbf{ r})=Z^{3/2} R_{nl} (Zr)Y_{lm} (\mathbf{ n}),
\end{equation}
where $Z$ is the charge of the nucleus, $n$ is the principal quantum number,
$l$ is the orbital angular momentum, $m$ is its projection, and $Y_{lm} (\mathbf{ n})$ is the spherical function.
The atomic system of units is used, where the electron mass $m_e=1$ and  the Bohr radius $a_{0} =1/(\alpha m)=1$.
$R_{nl} (Zr)$ satisfies the radial Schr\"odinger equation. The normalization condition of the radial part takes the form
\begin{equation}
Z^{3} \int_0^{\infty} r^{2} drR_{nl}^{2} (Zr) =1.
\end{equation}
This equation holds for any $Z$. Differentiating both parts of the equation by $Z$, we find
\begin{equation}
\int_0^{\infty} r^{3} drR_{nl} (Zr)R^{\prime}_{nl} (Zr) =-\frac{3}{2Z^{4} } .               \label{1111}
\end{equation}
The overlap amplitude of electron levels (OAEL) with the identical quantum numbers for atoms with nuclear charges $Z$ and $Z'=Z+\Delta Z$ can be written as follows:
\begin{equation} \label{exact}
O_{nl} = \int_0^{\infty} r^{2} drZ^{3/2} R_{nl} (Zr)Z^{\prime 3/2} R_{nl} (Z^{\prime}r).
\end{equation}
The overlap amplitude $O_{nl}$ determines the probability of finding the electron in its initial state after the decay.
We confine ourselves to the case of $Z \gg 1$, which covers medium-heavy and heavy atoms of the experimental interest.
Decomposing the term $Z^{\prime 3/2} R_{nl} (Z^{\prime}r)$ into a power series of $\Delta Z/Z$
to the second order, one obtains
\begin{eqnarray} \label{approx}
O_{nl} &=& \int_0^{\infty} r^{2} drZ^{3/2} R_{nl} (Zr)\\
&\times&
\left(1 +\Delta Z\frac{\partial}{\partial Z} +\frac{1}{2} \Delta Z^2\frac{\partial^{2} }{\partial Z^{2} } + \ldots \right) Z^{3/2} R_{nl} (Zr).  \nonumber
\end{eqnarray}
The second-order derivative term can be removed using the radial Schr\"odinger equation, the normalization condition and Eq.~(\ref{1111}).
The first-order derivative term vanishes because of condition (\ref{1111}). The first term in parentheses gives the normalization.
As a result, we obtain
\begin{eqnarray*}
O_{nl}&=& 1 + \frac{1}{2}  \left( -\frac{3}{4} +l(l+1)-2Z\left\langle r\right\rangle  \right.        \\
&~&~~~~~~~~~~+ \left. \frac{1}{n^{2} }Z^2 \left\langle r^{2} \right \rangle \right) \frac{\Delta Z^2}{Z^2} + \ldots,
\end{eqnarray*}
where the average radii are given by (see, e.g., \cite{LLQM})
\begin{eqnarray*}
Z\left \langle r \right\rangle  &=&  \frac{3n^{2} -l(l+1)}{2} , \\
Z^2\left\langle r^{2} \right\rangle &=& n^{2} \frac{5n^{2} +1-3l(l+1)}{2} .
\end{eqnarray*}

\begin{figure}
\vspace{-5mm}
\begin{center}
\resizebox{0.45\textwidth}{!}{%
  \includegraphics{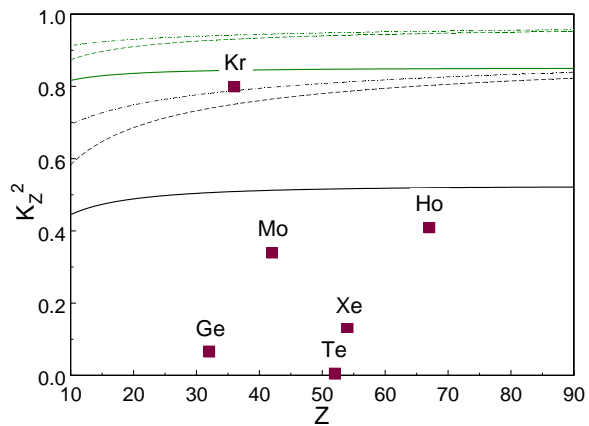}
}
\vspace{5mm}       
\caption{(color online)
Survival probability $K_Z^2$ of the electron shells as a function of the nuclear charge $Z$.
The green upper curves
correspond to the $ \beta$ decay processes with $\Delta Z = \pm 1$.
The black lower curves correspond to the double-$\beta$ decay processes with $\Delta Z = \pm 2$.
Dashed and dashed-dotted curves are calculated using Eqs.~(\ref{overlapshell}) and
(\ref{OTA}), respectively.
The solid curves are calculated as described in Sect.~2.2.
The dark red squares indicate the survival probability
$(K_{Z}^\mathrm{core~shells})^2$ of core electrons
in $\beta^-$ decay of Kr, electron capture in Ho, and $2\beta^-$ decays of Ge, Mo, Te and Xe,
calculated in Sects.~2.3 and 3.
}
\label{fig1}       
\end{center}
\end{figure}

Finally, we obtain
\begin{equation} \label{overlevel}
O_{nl} = 1-\frac{1}{8} \left(1+2n^{2} -2l(l+1)\right) \frac{\Delta Z^2}{Z^2} +...
\end{equation}
The condition $O_{nl}^2 \leq 1$ and the continuity in $Z$ are compatible only with the quadratic dependence in $\Delta Z$
and the negative second derivative of $O_{nl}$.

The OAELs in EC with $_{67}\mathrm{Ho}$ for the states $n=1,2,3,$ and $4$ and $l=0$
equal $O_{nl}= 0.999916$, $0.999749$, $0.999470$, and $0.999081$, whereas relativistic calculations based on the Dirac-Fock code of Ref.~\cite{AmandFaessler2015} yield $0.999910$, $0.999716$, $0.999389$, and $0.999332$, respectively. The OAELs for $n=2,3,$ and $4$ and $l=1 $ are $O_{nl}=0.999860$, $0.999582$, and $ 0.999192$, whereas Ref. \cite{AmandFaessler2015}
obtains $0.999801$, $0.999563$, and $0.999524$, respectively. The variance in the estimates does not exceed $3\cdot10^{-4}$.

We first consider atoms with closed shells. For this case,
the following relationship between the nuclear charge and the principal quantum number, $n_{Z} $, of the outermost
completely filled shell is given by
\begin{equation} \label{Zn}
Z = \sum _{n=1}^{n_{Z} } \sum _{l=0}^{n-1} \sum _{m=-l}^{l} \sum _{\sigma} 1 =\frac{n_{Z} (n_{Z} +1)(2n_{Z} +1)}{3},
\end{equation}
where the summation is carried out over the spin projection $\sigma = \pm 1/2$, the angular momentum projection $m$, the angular momentum $l$, and the principal quantum number $n$.

The overlap amplitude of the electron shells can be found by multiplying the OAELs of all the occupied levels by the principal quantum numbers $\le n_{Z} $:
\begin{equation} \label{overlap}
K_Z =\prod _{n=1}^{n_{Z} } \prod _{l=0}^{n-1}   \prod_{m=-l}^{l} \prod_{\sigma}   O_{nl},
\end{equation}
where the products account for the electron configuration of the electron shells, whereas the shielding of the nucleus by surrounding electrons is neglected.
In the EC and 2EC processes, one or two vacancies are formed in the electron shells.
Since the $O_{nl}$ are very close to unity, the levels corresponding to these vacancies need not be excluded from (\ref{overlap}).
The product can be evaluated with the help of the fact that for small $\epsilon$,
\begin{equation}
\prod_{k}(1+c_{k}\epsilon) \approx \exp\left(\sum_{k}c_{k}\epsilon\right), \label{prodexp}
\end{equation}
which yields
\begin{equation}
K_Z \approx \exp\left( -\frac{3(n_{Z}^{2} +n_{Z} +3)}{40 Z} \Delta Z^2 \right). \label{overlapshell}
\end{equation}
Equations (\ref{Zn}) - (\ref{overlapshell}) are derived for $Z$ corresponding to integer $n_Z$.
We analytically extend the overlap amplitude to arbitrary $Z$. In the limit of large $Z$,
\begin{eqnarray}
K_Z \approx  \exp\left(-{\frac {3^{5/3} 2^{1/3}}{80}} \frac {\Delta Z^2}{Z^{1/3}} \right).
\label{OTA}
\end{eqnarray}
Figure~\ref{fig1} shows $K_Z$ defined by Eqs.~(\ref{overlapshell}) and (\ref{OTA}) as a function of $Z$.
The approximation of Eq.~(\ref{overlapshell}) by Eq.~(\ref{OTA}) holds with an accuracy better than 3\%
for $Z \ge 10$ and $\Delta Z = \pm 1$  and better than 10\%
for $Z \ge 10$ and $\Delta Z = \pm 2$.

\begin{figure} %
\begin{center}
\resizebox{0.45\textwidth}{!}{%
  \includegraphics{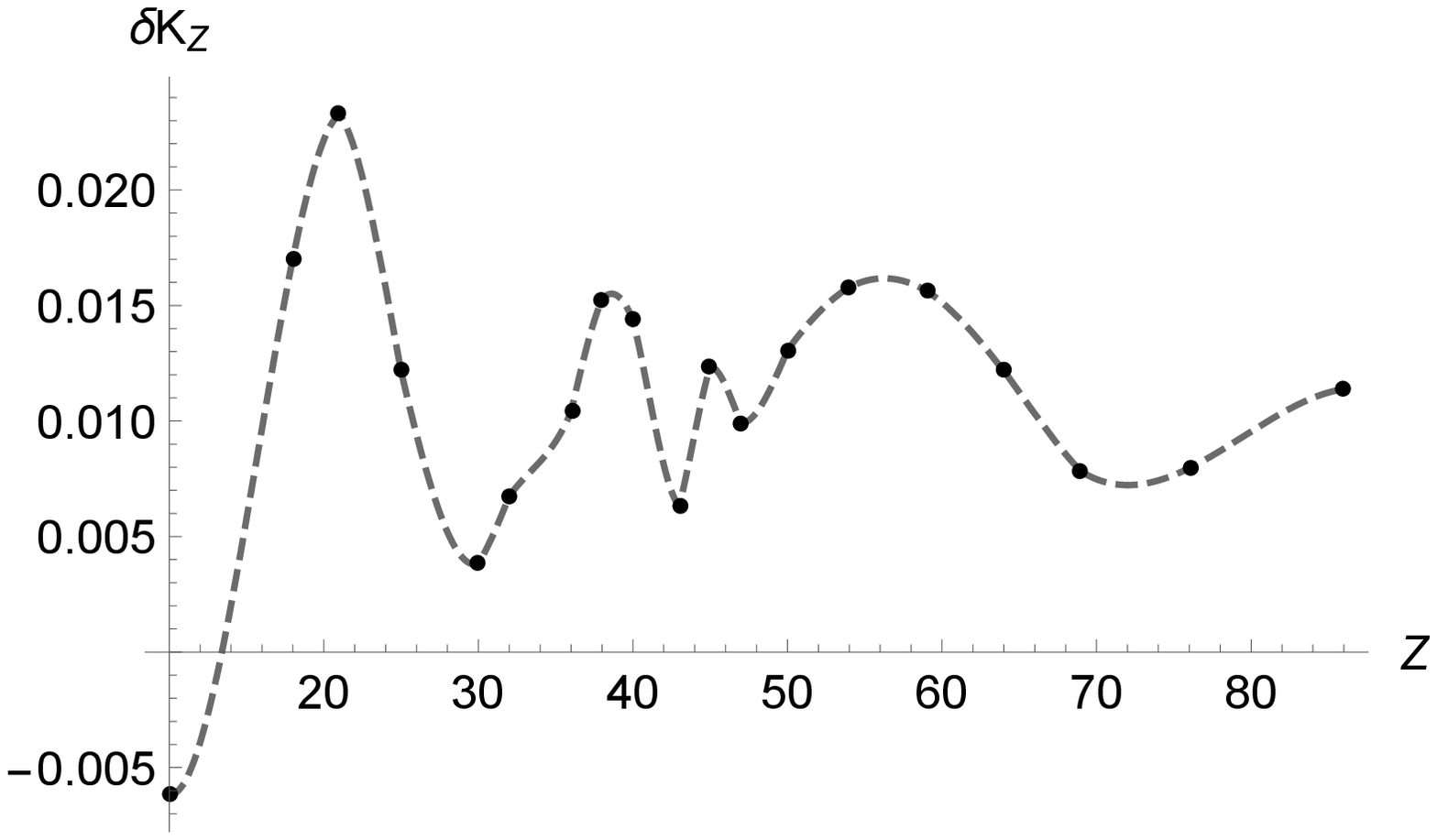}
}
\caption{
Accuracy of the $1/Z$ expansion used to derive Eq.~(\ref{OTA}).
The interpolating dashed curve passes through a finite set of values
$\delta K_{Z} = \Delta K_{Z}/K_{Z}$ calculated for $\beta$ decays of atoms with nuclear charge $10 \leq Z \leq 90$.
}
\label{fig3}
\end{center}
\end{figure}

Equation (\ref{OTA}) implies that EC in $^{163}$Ho is not accompanied by excitation of the electron shells
with the probability $K_Z^2 \approx 0.95$. In $2\beta^-$ decay of $^{76}$Ge, the survival probability equals $K_Z^2 \approx  0.75$.

The neutrinoless double-$\beta$ decay rate to channels with a low number of holes in the electron shells of the daughter atom
is proportional to
\begin{equation} \label{T12}
\Gamma^{0\nu 2\beta} \propto \left|K_Z m_{\beta\beta} g_A^2 \mathcal{M}^{0\nu 2 \beta} \right|^{2},
\end{equation}
where $\mathcal{M}^{0\nu 2 \beta} $ is the nuclear matrix element.
A similar dependence on $K_Z$ exists in the two-neutrino modes of $2\beta^{\pm}$ decays, $\Gamma^{2\nu 2\beta^{\pm}} \propto K_Z^{2}$,
and in $\beta$ decays, $\Gamma^{\beta} \propto K_Z^{2}$.
The overlap effect suppresses the decay rates and increases the upper limit on the neutrino mass $|m_{\beta\beta}|$,
determined from the neutrinoless $2\beta^-$ decay experiments \cite{Agostini:2018tnm,NEMO32015,CUORE2015,KamLAND-Zen:2016pfg}.

The accuracy of the $1/Z$ expansion can be tested by comparing
the overlap amplitude of the electron shells calculated with the use of
the exact OAELs (\ref{exact}) and the exact product (\ref{overlap}), on the one hand,
and
the approximate OAELs (\ref{approx}) and the approximate Eq.~(\ref{prodexp}) used to arrive at Eq.~(\ref{overlapshell}) on the other hand.
The $K_{Z}$ of atoms are found by multiplying
the OAELs (\ref{exact}) with powers corresponding to the number of electrons on each sub-shell;
the result is compared with Eq.~(\ref{overlapshell}).
The relative errors $\delta K_{Z} \equiv \Delta K_{Z} / K_{Z}$ are plotted in Fig.~\ref{fig3}.
The accuracy of the $1/Z$ expansion is better than 3\% for $Z \ge 10$ and $\Delta Z = \pm 1$.

\subsection{Analytic approach with shielding}

A more accurate estimate of the overlap amplitude of the electron shells can be obtained by taking into account the shielding
of the nucleus charge by electrons. Electrons in an atom move in an effective potential,
which can be approximated by the Coulomb potential with an effective charge $Z_ {\mathrm{eff}} < Z$.
An electron with a principal quantum number $n$ feels an effective charge $Z_{\mathrm{eff}} \approx Z - Z_s$,
where $Z_s$ is the number of electrons in the lower orbits with principal quantum numbers $1...n-1$.
$Z_s$ is given by Eq.~(\ref{Zn}) with $n_Z$ replaced by $n-1$.
Accordingly, in Eq.~(\ref{overlevel}) it is sufficient to make the substitution $Z \to Z_{\mathrm{eff}}$.
The overlap amplitude (\ref{overlap}) with the shielding effect taken into account
is computed in terms of special functions; however, the final expression is quite cumbersome,
so we focus on numerical estimates.
The solid curves in Fig.~\ref{fig1} show the total survival probabilities for $\Delta Z = \pm 1$
(upper curves) and $\pm 2$ (lower curves)
as functions of the nuclear charge $Z$; the estimates (\ref{overlapshell}) and (\ref{OTA}) are also shown.

Taking into account shielding, the OAELs in EC with $_{67}$Ho for $n = 1,2,3,$ and $4$ and $l = 0$ equal
$O_{nl} = 0.999916$, $0.999733$, $0.999269$, and $0.997287$, respectively,
while for $n = 2,3,$ and $4$ and $l = 1$, the OAELs equal $O_{nl} = 0.999852$, $0.999422$, and $0.997616$, respectively.
The difference from the calculations of Ref.~\cite{AmandFaessler2015}
is below $10^{-4}$ for $n \leq 3$ and $2\cdot10^{-3}$ for $n = 4$.

The OAEL of Eq.~(\ref{overlevel}) decreases with increasing $n$. The partial product
over the spin projection, the orbital angular momentum projection, and the orbital momentum yields
\begin{equation} \label{nnnn}
\prod_{l=0}^{n-1} \prod_{m = -l}^{l}\prod_{\sigma}O_{nl} \approx \exp\left( - n^2(n^2 + 2)\frac{\Delta Z^2}{Z^2} \right).
\end{equation}
The overlap amplitude of the single electron shell with the principal quantum number $n$
is determined by the ratio $\sim n^4/Z^2$.
Shielding of the nuclear charge is also important.
If $Z_{\mathrm{eff}}$ is a screened charge for electrons with principal quantum number $n$,
then the contribution to the overlap amplitude is enhanced as $\sim n^4/Z_{\mathrm{eff}}^2 >  n^4/Z^2$.
One can show that the error in $K_{Z}$ associated with the large-$Z$ approximation is below 2\%
starting from $Z = 10$ for $ \Delta Z = \pm 1 $.

Equations (\ref{overlevel}) and (\ref{nnnn}) show that
the maximum contribution to the suppression is given by electrons
with high values of $n$ and low values of $l$. The outermost electron shell
contribution is particularly important.

\begin{table*}[t]
\caption{Overlap amplitudes $O_{njl}$ of the electron levels with quantum numbers $n$, $j$, and $l$
for
$ \beta^- $ decay of $^{87}$Kr, electron capture in $^{163}$Ho, and $2\beta^-$ decays of $^{76}$Ge,
$^{100}$Mo, $^{130}$Te, and $^{136}$Xe
The electron binding energies $\epsilon^*$ from Ref.~\cite{Larkins1977} are given in keV for solid systems referenced to the Fermi level,
except for Kr and Xe, where $\epsilon^*$ are given for vapor-phase systems referenced to the vacuum level.
$Z_{\mathrm{eff}}$ is the effective nuclear charge determined from the Dirac equation.}
\label{TABX}
\centering
\scriptsize
\addtolength{\tabcolsep}{-1pt}

\begin{tabular}{|c|c|c|c|c|c|c|c|c|c|}
\hline\hline
$n2jl$ & $%
\begin{array}{c}
_{32}\mathrm{Ge} \\ \hline
\begin{array}{cc}
\epsilon ^{\ast }& \;\;\;\; Z_{\mathrm{eff}}%
\end{array}%
\end{array}%
$ & $%
\begin{array}{c}
_{34}\mathrm{Se} \\ \hline
\begin{array}{cc}
\epsilon ^{\ast }  & \;\;\;\; Z_{\mathrm{eff}}%
\end{array}%
\end{array}%
$ & $O_{njl}$ & $%
\begin{array}{c}
_{36}\mathrm{Kr} \\ \hline
\begin{array}{cc}
\epsilon ^{\ast }& \;\;\;\; Z_{\mathrm{eff}}%
\end{array}%
\end{array}%
$ & $%
\begin{array}{c}
_{37}\mathrm{Rb} \\ \hline
\begin{array}{cc}
\epsilon ^{\ast }& \;\;\;\; Z_{\mathrm{eff}}%
\end{array}%
\end{array}%
$ & $O_{njl}$
& $%
\begin{array}{c}
_{42}\mathrm{Mo} \\ \hline
\begin{array}{cc}
\epsilon ^{\ast }& \;\;\;\; Z_{\mathrm{eff}}%
\end{array}%
\end{array}%
$ & $%
\begin{array}{c}
_{44}\mathrm{Ru} \\ \hline
\begin{array}{cc}
\epsilon ^{\ast }  & \;\;\;\; Z_{\mathrm{eff}}%
\end{array}%
\end{array}%
$ & $O_{njl}$ \\ \hline\hline
$%
\begin{array}{c}
110 \\
210 \\
211 \\
231 \\
310 \\
311 \\
331 \\
332 \\
352 \\
410 \\
411 \\
431 \\
432 \\
452 %
\end{array}%
$ & $%
\begin{array}{cc}
11.1031 & 28.41 \\
1.4143 & 20.32 \\
1.2478 & 19.09 \\
1.2167 & 18.90 \\
0.1800 & 10.90 \\
0.1279 & 9.19 \\
0.1208 & 8.93 \\
0.0287 & 4.36 \\
0.0287 & 4.36 \\
0.0050 & 2.42 \\
0.0023 & 1.64 \\
   &    \\
   &    \\
   &    \\
\end{array}%
$ & $%
\begin{array}{cc}
12.6578 & 30.31 \\
1.6539 & 21.96 \\
1.4762 & 20.76 \\
1.4358 & 20.53 \\
0.2315 & 12.36 \\
0.1682 & 10.54 \\
0.1619 & 10.34 \\
0.0567 & 6.12 \\
0.0567 & 6.12 \\
0.0120 & 3.76 \\
0.0056 & 2.57 \\
   &    \\
   &    \\
   &    \\
\end{array}%
$ & $%
\begin{array}{c}
0.99832 \\
0.99303 \\
0.99553 \\
0.99570 \\
0.96272 \\
0.96516 \\
0.96031 \\
0.90397 \\
0.90398 \\
0.36980 \\
0.41864 \\
        \\
        \\
        \\
\end{array}%
$ & $%
\begin{array}{cc}
14.3256&32.21\\
1.9210&23.65\\
1.7272&22.43\\
1.6749&22.17\\
0.2921&13.88\\
0.2218&12.10\\
0.2145&11.90\\
0.0950&7.92\\
0.0938&7.87\\
0.0275&5.68\\
0.0147&4.16\\
0.0140&4.06\\
   &    \\
   &    \\
\end{array}%
$ & $%
\begin{array}{cc}
15.1997&33.17\\
2.0651&24.51\\
1.8639&23.30\\
1.8044&23.01\\
0.3221&14.57\\
0.2474&12.78\\
0.2385&12.55\\
0.1118&8.60\\
0.1103&8.54\\
0.0293&5.87\\
0.0148&4.17\\
0.0140&4.06\\
   &    \\
   &    \\
\end{array}%
$ & $%
\begin{array}{c}
0.99966\\
0.99850\\
0.99908\\
0.99913\\
0.99431\\
0.99439\\
0.99473\\
0.99421\\
0.99427\\
0.99586\\
0.99996\\
1.00000 \\
        \\
        \\
\end{array}%
$ & $
\begin{array}{cc} 
19.9995 & 37.95 \\
2.8655 & 28.81 \\
2.6251 & 27.60 \\
2.5202 & 27.18 \\
0.5046 & 18.22 \\
0.4097 & 16.43 \\
0.3923 & 16.10 \\
0.2303 & 12.34 \\
0.2270 & 12.25 \\
0.0618 & 8.52 \\
0.0348 & 6.39 \\
0.0348 & 6.39 \\
0.0018 & 1.45 \\
0.0018 & 1.45 \\
\end{array}%
$ & $%
\begin{array}{cc}
22.1172 &39.87\\
3.2240 & 30.54\\
2.9669 & 29.31\\
2.8379 & 28.84\\
0.5850 & 19.62\\
0.4828 & 17.83\\
0.4606 & 17.44\\
0.2836 & 13.69\\
0.2794 & 13.59\\
0.0749 & 9.38\\
0.0431 & 7.12\\
0.0431 & 7.12\\
0.0020 & 1.54\\
0.0020 & 1.54\\
\end{array}%
$ & $%
\begin{array}{c}
0.99898\\
0.99599\\
0.99759\\
0.99778\\
0.98696\\
0.98732\\
0.98794\\
0.99055\\
0.99060\\
0.96224\\
0.95900\\
0.95902\\
0.99273\\
0.99273\\
\end{array}%
$
\\ \hline\hline
$n2jl$ & $%
\begin{array}{c}
_{52}\mathrm{Te}\\ \hline
\begin{array}{cc}
\epsilon ^{\ast }& \;\;\;\; Z_{\mathrm{eff}}%
\end{array}%
\end{array}%
$ & $%
\begin{array}{c}
_{54}\mathrm{Xe} \\ \hline
\begin{array}{cc}
\epsilon ^{\ast }  & \;\;\;\; Z_{\mathrm{eff}}%
\end{array}%
\end{array}%
$ & $O_{njl}$ & $%
\begin{array}{c}
_{54}\mathrm{Xe} \\ \hline
\begin{array}{cc}
\epsilon ^{\ast }& \;\;\;\; Z_{\mathrm{eff}}%
\end{array}%
\end{array}%
$ & $%
\begin{array}{c}
_{56}\mathrm{Ba} \\ \hline
\begin{array}{cc}
\epsilon ^{\ast }& \;\;\;\; Z_{\mathrm{eff}}%
\end{array}%
\end{array}%
$ & $O_{njl}$
& $%
\begin{array}{c}
_{67}\mathrm{Ho} \\ \hline
\begin{array}{cc}
\epsilon ^{\ast }& \;\;\;\; Z_{\mathrm{eff}}%
\end{array}%
\end{array}%
$ & $%
\begin{array}{c}
_{66}\mathrm{Dy} \\ \hline
\begin{array}{cc}
\epsilon ^{\ast }  & \;\;\;\; Z_{\mathrm{eff}}%
\end{array}%
\end{array}%
$ & $O_{njl}$ \\ \hline\hline
$%
\begin{array}{c}
110 \\
210 \\
211 \\
231 \\
310 \\
311 \\
331 \\
332 \\
352 \\
410 \\
411 \\
431 \\
432 \\
452 \\
453 \\
473 \\
510 \\
511 \\
531 %
\end{array}%
$ & $%
\begin{array}{cc} 
31.8138&47.60\\
4.9392&37.65\\
4.6120&36.41\\
4.3414&35.65 \\
1.0060&25.68\\
0.8697&23.89\\
0.8187&23.24\\
0.5825&19.61\\
0.5721&19.45 \\
0.1683&14.05\\
0.1102&11.38\\
0.1102&11.38\\
0.0398&6.84\\
0.0398&6.84\\
      &       \\
      &       \\
0.0116&4.61\\
0.0023&2.06\\
0.0023&2.06\\
\end{array}%
$ & $%
\begin{array}{cc} 
34.5644 & 49.54 \\
5.4528 & 39.51 \\
5.1037 & 38.25 \\
4.7822 & 37.41 \\
1.1487 & 27.43 \\
1.0021 & 25.63 \\
0.9406 & 24.91 \\
0.6894 & 21.33 \\
0.6767 & 21.15 \\
0.2133 & 15.82 \\
0.1455 & 13.07 \\
0.1455 & 13.08 \\
0.0695 & 9.04  \\
0.0675 & 8.91  \\
&       \\
&       \\
0.0234&6.56\\
0.0134&4.96\\
0.0121&4.71\\
\end{array}%
$ & $%
\begin{array}{r}
0.99928\\
0.99712\\
0.99831\\
0.99852\\
0.98940\\
0.99046\\
0.99095\\
0.99378\\
0.99384\\
0.94280\\
0.93124\\
0.93138\\
0.80895\\
0.82734\\
  \\
  \\
0.36726\\
-0.34644\\
-0.36107
\end{array}%
$ & $%
\begin{array}{cc} 
34.5644 & 49.54 \\
5.4528 & 39.51 \\
5.1037 & 38.25 \\
4.7822 & 37.41 \\
1.1487 & 27.43 \\
1.0021 & 25.63 \\
0.9406 & 24.91 \\
0.6894 & 21.33 \\
0.6767 & 21.15 \\
0.2133 & 15.82 \\
0.1455 & 13.07 \\
0.1455 & 13.08 \\
0.0695 & 9.04  \\
0.0675 & 8.91  \\
      &       \\
      &       \\
0.0234 & 6.56  \\
0.0134 & 4.96  \\
0.0121 & 4.72  \\
\end{array}%
$ & $%
\begin{array}{cc}
37.4406 & 51.49 \\
5.9888 & 41.35 \\
5.6236 & 40.10 \\
5.2470 & 39.17 \\
1.2928 & 29.08 \\
1.1367 & 27.28 \\
1.0622 & 26.47 \\
0.7961 & 22.92 \\
0.7807 & 22.72 \\
0.2530 & 17.22 \\
0.1918 & 15.00 \\
0.1797 & 14.53 \\
0.0925 & 10.43  \\
0.0899 & 10.28  \\
      &       \\
      &       \\
0.0291 & 7.31  \\
0.0166 & 5.52  \\
0.0146 & 5.18  \\
\end{array}%
$ & $%
\begin{array}{r}
0.99933 \\
0.99740 \\
0.99845 \\
0.99863 \\
0.99156 \\
0.99243 \\
0.99304 \\
0.99546 \\
0.99553 \\
0.97004 \\
0.93194 \\
0.96003 \\
0.94724 \\
0.94702 \\
  \\
  \\
0.92576 \\
0.93386 \\
0.94892
\end{array}%
$ & $%
\begin{array}{cc} 
55.6177 & 62.17 \\
9.3942 & 51.35 \\
8.9178 & 50.09 \\
8.0711 & 48.52 \\
2.1283 & 37.17 \\
1.9228 & 35.36 \\
1.7412 & 33.85 \\
1.3915 & 30.28 \\
1.3514 & 29.88 \\
0.4357 & 22.57 \\
0.3435 & 20.05 \\
0.3066 & 18.97 \\
0.1610 & 13.75 \\
0.1610 & 13.76 \\
0.0037 & 2.09 \\
0.0037 & 2.09 \\
0.0512 & 9.70 \\
0.0203 & 6.11 \\
0.0203 & 6.11 \\
\end{array}%
$ & $%
\begin{array}{cc}
53.7885 & 61.20 \\
9.0458 & 50.43 \\
8.5806 & 49.17 \\
7.7901 & 47.67 \\
2.0468 & 36.46 \\
1.8418 & 34.62 \\
1.6756 & 33.21 \\
1.3325 & 29.63 \\
1.2949 & 29.25 \\
0.4163 & 22.07 \\
0.3318 & 19.71 \\
0.2929 & 18.55 \\
0.1542 & 13.46 \\
0.1542 & 13.46 \\
0.0042 & 2.22 \\
0.0042 & 2.22 \\
0.0629 & 10.75 \\
0.0263 & 6.95 \\
0.0263 & 6.95 \\
\end{array}%
$ & $%
\begin{array}{c}
0.99987 \\
0.99956 \\
0.99975 \\
0.99980 \\
0.99906 \\
0.99910 \\
0.99930 \\
0.99959 \\
0.99960 \\
0.99784 \\
0.99890 \\
0.99810 \\
0.99878 \\
0.99878 \\
0.99549 \\
0.99549 \\
0.93362 \\
0.90411 \\
0.90415 \\
\end{array}%
$
\\ \hline\hline
\end{tabular}
\end{table*}

\subsection{Relativistic approach}

Since the OAELs are close to unity ($O_{nl} - 1 \sim 10^{-2} \div 10^{-4}$), when
using software packages for modeling atomic systems, the accuracy of the numerical determination
of the electron wave functions should be monitored closely.
The use of analytical expressions for electron wave functions has certain advantages in this respect,
as analytical methods narrow the range of uncertainties inherent in numerical schemes.
A fairly accurate treatment of the overlap effect can be made on the basis
of the relativistic Dirac equation in a Coulomb field  by determining
the effective nuclear charge for each electron level from the known semi-empirical values of the electron binding energies \cite{Larkins1977}.
This approach was used earlier to calculate the Coulomb interaction energy of electron
holes for the neutrinoless 2EC problem \cite{KRIV11}.
Using the known electron binding energies,
the effective charges of the parent and daughter nuclei are calculated for each electron level and then substituted into
the relativistic energy eigenfunctions of electrons in the Coulomb field. The OAEL
\begin{eqnarray}
O_{njl} &=& \int_0^{\infty} r^{2} dr(f_{njl}(Z,r)f_{njl}(Z^{\prime},r) \nonumber \\
           &&~~~~~~~\;\;~+  g_{njl}(Z,r)g_{njl}(Z^{\prime},r)), \label{Onjl}
\end{eqnarray}
with $f_{njl}$ and $g_{njl}$ being the radial functions of bispinor components,
is calculated numerically with MAPLE \footnote{$\mathrm{https://www.maplesoft.com/}$} without resorting to a large-$Z$ expansion.

\section{Numerical results and discussion}

The OAELs (\ref{Onjl}) for $\beta^-$ decay of Kr, electron capture in Ho, and $2\beta^-$ decays of Ge, Mo, Te, and Xe are presented in Table~1; the corresponding electron energies and the screened effective nuclear charges are also given.

The estimates presented in Table~1 demonstrate the dominance of contributions
of high electron levels and in particular valence electrons.
Since valence electrons participate in bonding and collectivize in metals, their effect is the least controlled.
A straightforward calculation covering the valence shell contributions
yields rather irregular values of the total overlap amplitude $K_Z$ reported in Table~2.
The overlap amplitudes of the electron shells in Kr and Xe decays 
neglect the difference between the solid and vapor-phase systems, for which the electron binding energies are reported in Ref.~\cite{Larkins1977}.
The energy of the $5s$ electron level in Mo is not given in Ref.~\cite{Larkins1977}, so this level is not included in $K_Z$.
The effects of electron holes formed in the core shells of the daughter atom in EC or 2EC processes
and of missing electrons on the valence shell of the daughter atom in $\beta^-$ and $2\beta^-$ decays
are disregarded because each individual level has only a weak impact on $K_Z$.
Low values of $K_Z$ imply that the decay mode in which electron shells of the daughter atom are unexcited
is not dominant. This obviously applies to the decays of germanium and tellurium.
To judge which channels dominate in cases with low $K_Z$, it is necessary to consider transitions to excited states.

The survival probability of core electrons is less affected by the environment.
Excluding 4 valence electrons in Ge, we obtain
$K_{Z}^{\mathrm{core~shells}} = 0.26$,
which should be compared to $K_Z =  6.2\cdot10^{-3} \ll K_{Z}^{\mathrm{core~shells}}$.
A similar calculation for
Kr without 8,
Mo without 6,
Te without 6,
Xe without 8, and
Ho without 3
valence electrons
yields the values reported in Table~2.
To estimate uncertainties in $K_Z^{\mathrm{core~shells}}$, it is necessary to perform
the calculations using alternative schemes for modeling the electron shell structure.
We remark that the outermost subshells of krypton and
xenon are filled.
Under standard conditions for temperature and pressure, krypton and xenon are inert noble gases
that do not usually form chemical bonds with other atoms.
The number of valence electrons is, however, the average characteristic of atoms.
For example,
holmium has a maximum valence of 13, while in chemical compounds, it usually donates three electrons.
Krypton usually donates two electrons while forming compounds, etc.
The ratio $K_Z/K_Z^{\mathrm{core~shells}}$ determines the contribution of the valence shell.
The strong inequality $K_{Z} \ll K_{Z}^{\mathrm{core~shells}}$, which characterizes germanium and tellurium decays,
implies that their dominant decay channels are associated with excitations of the valence shells.
Given that the overlap amplitude of the electron shells depends on the second power of $ \Delta Z$, according to Eq.~(\ref{OTA}),
one can expect similar results for the EC process in $^{81}$Kr, discussed in Ref.~\cite{Ratkevich2017},
and 2EC process in $^{124}$Xe, which has recently been observed by the XENON Collaboration~\cite{XENON2019}.

The amplitudes of neutrinoless $2\beta^-$ decay and neutrinoless 2EC processes are proportional
to the effective electron neutrino Majorana mass
\begin{equation}
m_{\beta\beta} = \sum_i U_{ei}^2 m_i,
\end{equation}
where $U_{\alpha i}$ is the Pontecorvo--Maki--Nakagawa--Sakata mixing matrix and $m_i$ are the
diagonal neutrino masses.
The study of cosmic microwave background anisotropies by the Planck Collaboration yields $\sum_i m_i < 0.12$ eV \cite{Aghanim2018}.

Exotic interactions beyond the Standard Model can modify the mass of neutrinos in nuclear matter,
so the effective electron neutrino Majorana mass observed in $2\beta^-$ decays and 2EC processes may
differ from the vacuum value \cite{Kovalenko2014}.
A similar effect arises from supersymmetric generalizations of the Standard Model \cite{Mohapatra1986,Vergados1987}.

The GERDA and KamLAND-Zen collaborations \cite{Agostini:2018tnm,KamLAND-Zen:2016pfg}, from searching the neutrinoless $2\beta^-$ decay with $^{76}$Ge and $^{136}$Xe, give restrictions on the effective electron neutrino Majorana mass,
$|m_{\beta\beta}| <  120 - 260$ meV and $ 61 - 165 $ meV, respectively,
by taking into account uncertainties in the nuclear matrix elements and for the unquenched axial-vector coupling $g_A = 1.27$.
The constraints $|m_{\beta\beta}| <  330 - 620$ meV and $270 - 760$ meV are obtained by the NEMO-3 and CUORE collaborations \cite{NEMO32015,CUORE2015}
from searching the neutrinoless $2\beta^-$ decay with $^{100}$Mo and $^{130}$Te, respectively.

The neutrino mass limits must be supplemented by a new condition, $K_Z = 1$, according to relation (\ref{T12}).
The imperfect overlap of the electron shells with moderately low $K_Z$ leads to an increase in the upper limits.
For neutrinoless $ 2 \beta^-$ decays of $^{100} $Mo and $ ^{136}$Xe,
the neutrino mass limits increase by $1 / K_Z = $ 1.8 and 4.6 times, respectively,
and the environment-independent part of the scaling factors equals $1/$ $K_Z^{\mathrm{core~shells}}=$ 1.7 and 2.8, respectively.

The low ratios $K_Z / K_Z^{\mathrm{core~shells}}$ signal the importance of decay channels with excited valence shells of the daughter atoms.
The low values of $ K_Z^{\mathrm{core~shell}}$ further indicate a noticeable contribution of channels with excited core electrons.
The obviously small overlap, $K_Z \ll 1$, observed for the neutrinoless $ 2\beta^- $ decays of $ ^{76} $Ge and $ ^{130} $Te
signifies dominance of decay channels with multiple excited electron shells,
which takes us beyond the applicability of relationship (\ref{T12}).
In the neutrinoless $ 2\beta^- $ decays of $^{100}$Mo and $ ^{136}$Xe,
$K_Z$  and $ K_Z^{\mathrm{core~shells}}$ deviate from unity only moderately;
the probability of such decays obeys relationship (\ref{T12}), which is
valid for the daughter atoms with a low number of holes in the electron shells.

\begin{table*}[t]
\caption{
Total overlap amplitude $K_{Z}$ and overlap amplitude of core electrons $K_{Z}^{\mathrm{core~shells}}$
in $ \beta^- $ decay of $^{87}$Kr, electron capture in $^{163}$Ho, and $2\beta^-$ decays of $^{76}$Ge,
$^{100}$Mo, $^{130}$Te, and $^{136}$Xe.
The number of valence electrons of the parent atoms is given in parentheses after the atom symbols.
The electron configuration of the parent atoms is shown below the atom symbols.}
\label{TABY}
\centering
\scriptsize
\addtolength{\tabcolsep}{-1pt}

\begin{tabular}{|c|c|c|c|c|c|c|}
\hline\hline
Overlap & $_{32}\mathrm{Ge}$~(4) & $_{36}\mathrm{Kr}$~(8) & $_{42}\mathrm{Mo}$~(6) & $_{52}\mathrm{Te}$~(6) & $_{54}\mathrm{Xe}$~(8) & $_{67}\mathrm{Ho}$~(3) \\
amplitude &  [Ar]$ 3d^{10} 4s^2 4p^2$ & [Ar]$ 3d^{10} 4s^2 4p^6$ & [Kr]$ 4d^5 5s^1$& [Kr]$ 4d^{10} 5s^2 5p^4$ & [Kr]$ 4d^{10} 5s^2 5p^6$ & [Xe]$ 4f^{11} 6s^2$\\
\hline\hline
$K_{Z}$ & $6.2\cdot 10^{-3}$ & 0.89 & 0.56 & $1.4\cdot 10^{-4}$ & 0.22 & 0.53 \\
$K_{Z}^{\mathrm{core~shells}}$ & 0.26 & 0.90 & 0.58 & 0.069 & 0.36 & 0.64 \\
\hline\hline
\end{tabular}
\end{table*}

The normalized energy spectrum of electron capture in atoms carries information about the electron neutrino mass \cite{Gastaldo:2013wha,Hassel:2016ixd}.
In Ref.~\cite{Brass2018}, \textit{ab-initio} calculation of excited states up to three electron holes, associated with EC in $^{163}$Ho,
is performed for transitions with branching ratios above $10^{-3}$.
We remark that $K_Z$ cancels out from and does not influence the normalized energy spectra 
of $\beta$ and double-$\beta$ decays
to channels with a low number of holes in the electron shells of the daughter atoms.

Calorimetric detectors for $\beta$ and double-$ \beta $ decays measure the total energy released.
Suppression of channels with a small number of holes in the electron shells
can be compensated by channels associated with multiple excitation of the electron shells of the daughter atoms.
Since the binding energy of valence electrons in an atom is approximately 10 eV, to experimentally distinguish
the contribution of channels with a small number of holes in the electron shells
from the contribution of channels with multiple excitation of the electron shells, it is required to measure
the energy of electrons emitted in $ \beta^-$ and $ 2\beta^-$ decays with a resolution of 10 eV or better.
If this strong condition is not satisfied, then the channels in question cannot be distinguished experimentally.

The overlap amplitude of valence shells, $K_Z/K_Z^{\mathrm{core~shells}}$,
is sensitive to chemical bonds, properties of the environment, etc.;
these uncertainties are comparable to the uncertainties inherent
in a nuclear structure calculation.
Uncertainties in nuclear matrix elements of neutrinoless $2\beta^-$ decay reach 50\%
(see, e.g., \cite{Simkovic:2007vu}).
The axial-vector coupling in nuclei is also quite uncertain;
it can be half of its value for a free nucleon \cite{Suhonen:2017krv},
which can in turn increase the upper limit of $|m_{\beta\beta}|$ by approximately 4 times.
In the case of neutrinoless $ 2 \beta^- $ decay of $ ^{136}$Xe,
the valence shell contribution increases the upper limit
by $K_Z^{\mathrm{core~shells}}/K_Z = 1.6$ times. In the decay of $^{100}$Mo,
the effect is negligible,
while in the decays of $ ^{76} $Ge and $ ^{130} $Te with $K_Z \ll 1$, the valence shell contribution
can hardly be quantified with regard to $|m_{\beta\beta}|$ because of
the involvement of multiple excited states of electron shells of the daughter atoms.

The experimental limits for neutrinoless 2EC half-lives are weaker compared to neutrinoless $2\beta^-$ decay.
One of the best such limits is obtained for the $^{40}$Ca $\to ^{40}$Ar decay:
$T_{1/2}^{0\nu\mathrm{2EC}} > 1.4\cdot 10^{22}$ years \cite{Angloher2016}.
The articular interest in the neutrinoless 2EC is associated with the possibility of resonance enhancement of the decay rate \cite{KRIV11,Eliseev:2011zza}.
The valence shell excitations have a strong effect on the $K_Z$ value but are not associated with absorption of a significant amount of energy.
The typical binding energy of valence electrons is 10 eV. The resonance condition for
the neutrinoless 2EC process is determined by the natural width of the electron levels, which has a similar order of magnitude.
Transitions accompanied by the excitations of valence electrons do not strongly violate the resonance condition
but can increase $K_Z$. Holmium atoms used to study the single EC
are embedded in a metallic environment, where the collectivization of valence electrons
complicates the estimate of $K_Z$. The same observation applies to any EC experiments
using a substance in a metallic phase.

\section{Conclusions}

In this paper, the overlap amplitude of electron shells in $ \beta$ and double-$\beta $ decays was considered.
The deviation of OAELs from unity increases with an increase in the principal quantum number and a decrease in the orbital angular momentum.
A simple estimate (\ref{OTA}) shows that the product of OAELs significantly deviates from unity
despite the fact that the values of the OAELs are very close to unity.
The total overlap amplitude $K_Z$ is found to be sensitive to the electron configuration of the outer shells.
The valence electrons provide a particularly large but irregular contribution to $K_Z$.
Given that valence electrons participate in bonding and collectivize in metals,
the phase state of the substance, impurities, and molecular structure of the elements affect the value of $K_Z$.
The overlap amplitude of core electrons, $K_Z^{\mathrm{core~shells}}$, is less sensitive to the environment.
The relativistic self-consistent scheme described in Sect.~2.3 was used to calculate $K_Z$ and $K_Z^{\mathrm{core~shells}}$
for $ \beta^- $ decay of $^{87}$Kr, electron capture in $^{163}$Ho, and $2\beta^-$ decays of $^{76}$Ge,
$^{100}$Mo, $^{130}$Te, and $^{136}$Xe.
The results show that the overlap effect is numerically significant or highly significant, with $K_Z^{\mathrm{core~shells}}$
ranging from 0.90 (Kr) to 0.069 (Te). The overlap amplitude of core electrons provides an upper limit
for the total overlap amplitude, which depends on the environment.

In the dominant channels of $ \beta$ and double-$\beta $ decays,
the electron shells of the daughter atoms, as hitherto assumed, remain with a low number of excitations.
In transitions of this type, the overlap amplitude $K_Z$ cancels out from
the normalized energy spectra and branching ratios, while the total decay rates are suppressed as $\Gamma^{\beta/2\beta} \sim K_Z^2$.
Such suppression requires renormalization of the nuclear matrix elements and
the axial-vector coupling $g_A$ extracted from $ \beta$ and double-$\beta $ decay experiments.
One can assume that the observed quenching of $g_A$ is partly
due to the imperfect overlap of the electron shells.

The molybdenum atom 
has particularly stable overlap factors in the $2\beta^-$ decay:
$K_{Z} = 0.56$ and $K_{Z}^{\mathrm{core~shells}} = 0.58$.
The neutrino mass limits extracted from the
experimental searches for neutrinoless $2\beta^-$ decay of $^{100}$Mo
suffer from only minimal uncertainties related to the overlap of the electron shells.
For krypton and xenon, we obtained $K_{Z} = 0.89$ and $K_{Z}^{\mathrm{core~shells}} = 0.90$
and obtained $K_{Z} = 0.22$ and $K_{Z}^{\mathrm{core~shells}} = 0.46$, respectively.
Since krypton and xenon have more stable configurations than other elements of the periodic table,
their overlap amplitudes are expected to be weakly dependent on the environment.
The overlap effect with moderately low $K_Z$ modifies the relationship between $m_ {\beta\beta}$ and the neutrinoless $2\beta$ decay widths
as described by Eq.~(\ref{T12}).
The overlap amplitudes of $^{76}$Ge and $^{130}$Te are surprisingly low and sensitive to the contribution
of the valence shells. Therefore, the dominant $2\beta^-$ decay channels of $^{76}$Ge and $^{130}$Te contain
daughter atoms with multiple excited electron shells. At $K_Z \ll 1$, the relationship between $m_ {\beta\beta}$ and
the neutrinoless $2\beta$ decay widths is more complex than that in Eq.~(\ref{T12}).

The overlap effect is important for fitting phenomenological parameters of nuclear structure models and
determination of quantitative relationships between the half-lives of neutrinoless $2\beta$ decays
and the effective electron neutrino Majorana mass.

\vspace{2mm}
The authors are indebted to F. Danevich for the discussion of experimental limitations in the $\beta$ spectrum measurements.
This work is partially supported by RFBR Grant No.~18-02-00733.

%
%

\end{document}